\documentstyle[12pt,amssymb,psfig]{article}
\newcommand{\be}{\begin{eqnarray}}
\newcommand{\ee}{\end{eqnarray}}

\renewcommand{\simeq}{\stackrel\sim=}

\newcommand{\myhead}{\sf J. Chu and C. Adami}
\begin{document}
%\thepage
\pagestyle{myheadings}
\markboth{\myhead}{\myhead}
\begin{titlepage}
\vskip 7cm
\centerline{\Large\bf A Simple Explanation for}
\centerline{\Large\bf Taxon Abundance Patterns}
\vskip 0.5in
\centerline{Johan Chu and Christoph Adami}
\vskip 0.25in
\centerline{\it W.K.\ Kellogg Radiation Laboratory 106-38}
\centerline{\it California Institute of Technology, Pasadena, CA 91125} 
\vskip 3.5 cm
\noindent{\it Classification:}\\
\noindent{\it  Biological Sciences (Evolution), Physical
  Sciences (Applied Mathematics)}
\vskip 2cm
\noindent Corresponding author:
\vskip 0.25in
Dr. Chris Adami

E-mail: adami@krl.caltech.edu

Phone: (+) 626 395-4256

Fax: (+) 626 564-8708

\end{titlepage}
\setcounter{page}{2}
\begin{abstract}
  For taxonomic levels higher than
  species, the abundance distributions of number of subtaxa per taxon
  tend to approximate power laws,  but often show strong deviationns
  from such a law.  Previously, these deviations were attributed
  to finite-time effects in a continuous time branching process at the
  generic level. Instead, we describe here a simple discrete branching
  process which generates the observed distributions and find that the
  distribution's deviation from power-law form is not caused by
  disequilibration, but rather that it is time-independent and
  determined by the evolutionary properties of the taxa of interest.
  Our model predicts---with no free parameters---the rank-frequency
  distribution of number of families in fossil marine animal orders
  obtained from the fossil record.  We find that near power-law
  distributions are statistically almost inevitable for taxa higher
  than species.  The branching model also sheds light on species
  abundance patterns, as well as on links between evolutionary
  processes, self-organized criticality and fractals.
\end{abstract}

Taxonomic abundance distributions have been studied since the
pioneering work of Yule~\cite{YULE}, who proposed a continuous time
branching process model to explain the distributions at the generic
level, and found that they were power laws in the limit of
equilibrated populations. Deviations from the geometric law were
attributed to a finite-time effect, namely, to the fact that the
populations had not reached equilibrium. Much later,
Burlando~\cite{BURLANDO1,BURLANDO2} compiled data that appeared to
corroborate the geometric nature of the distributions, even though
clear violations of the law are visible in his data also. In this
paper, we present a model which is based on a discrete branching process 
whose distributions are time-independent and where violations of the
geometric form reflect specific environmental conditions and pressures
that the assemblage under consideration was subject to during
evolution. As such, it holds the promise that an analysis of taxonomic
abundance distributions may reveal certain characteristics of
ecological niches long after its inhabitants have disappeared.

The model described here is based on the simplest of branching
processes, known in the mathematical literature as the {\it
  Galton-Watson} process. Consider an assemblage of taxa at one
taxonomic level. This assemblage can be all the families under a
particular order, all the subspecies of a particular species, or any
other group of taxa at the same taxonomic level that can be assumed to
have suffered the same evolutionary pressures. We are interested in
the shape of the rank-frequency distribution of this assemblage and
the factors that influence it.

We describe the model by explaining a specific example: the
distribution of the number of families within orders for a particular
phylum. The adaptation of this model to different levels in the
taxonomic hierarchy is obvious.  We can assume that the assemblage was
founded by one order in the phylum and that this order consisted of
one family which had one genus with one species.  We further assume
that new families in this order are created by means of mutation in
individuals of extant families.  This can be viewed as a process where
existing families can ``replicate'' and create new families of the
same order, which we term {\em daughters} of the initial family.  Of
course, relatively rarely, mutations may lead to the creation of a new
order, a new class, etc. We define a probability $p_i$ for a family to
have $i$ daughter families of the same order ({\em true daughters}).
Thus, a family will have no true daughters with probability $p_0$, one
true daughter with probability $p_1$, and so on.  For the sake of
simplicity, we initially assume that all families of this phylum share
the same $p_i$.  We show later that variance in $p_i$ among different
families does not significantly affect the results, in particular the
shape of the distribution. The branching process described above gives
rise to an abundance distribution of families within orders, and its
probability distribution can be obtained from the Lagrange expansion
of a nonlinear differential equation~\cite{HARRIS}.  Using a simple
iterative algorithm~\cite{JCCA2} in place of this Lagrange expansion
procedure, we can calculate rank-frequency curves for many different
sets of $p_i$.  It should be emphasized here that we are mostly
concerned with the shape of this curve for $n \lesssim 10^4$, and not
the asymptotic shape as $n \rightarrow \infty$, a limit that is not
reached in nature.

For different sets of $p_i$, the theoretical curve
can either be close to a power-law, a power law with an exponential tail
or a purely exponential distribution (Fig.\ 1). 
\begin{figure}[h]
\centerline{\psfig{figure=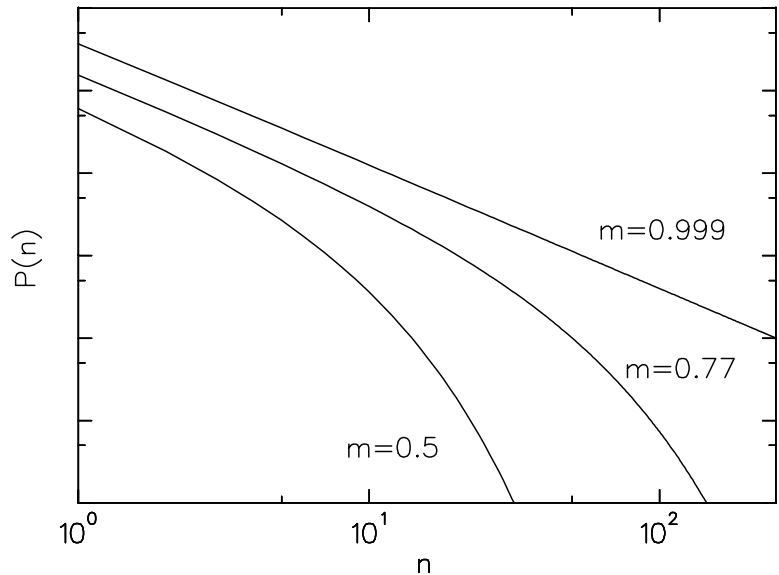,width=4in,angle=90}}
\caption{Predicted abundance pattern $P(n)$
(probability for a taxon to have $n$ subtaxa) of the
branching model with different values of $m$. The
curves have been individually rescaled. }
\end{figure}
We show here that there is a
global parameter that distinguishes among these cases. 
Indeed, the mean number of true daughters, i.e., the mean number
of different families of the same order that each family gives 
rise to in the example above,  
\be
m = \sum_{i=0}^{\infty} i \cdot p_i
\ee
is a good indicator of the overall shape of the curve.
Universally, $m = 1$ leads to a power law for the abundance distribution.
The further 
$m$ is away from $1$, the further the curve diverges from
a power-law and towards an exponential curve. The value
of $m$ for a particular assemblage can be estimated from
the fossil record also, allowing for a characterization
of the evolutionary process with no free parameters. Indeed,  
if we assume that the number of families in this phylum existing at one
time is roughly constant, or varies slowly compared to the average rate
of family creation 
(an assumption the fossil record seems to vindicate~\cite{RAUP}), 
we find that $m$ can be related to the ratio $R_o/R_f$
of the rates of creation of orders and families---by
\be
m = (1+\frac{R_o}{R_f})^{-1}
\ee
to leading order~\cite{JCCA2}.

In general, we can not expect all the families within an order to share
the same $m$. Interestingly, it turns out that 
even if the $p_i$ and $m$ differ
widely between different families, the rank-frequency curve is 
identical to that obtained by assuming a fixed $m$ equal to the average
of $m$ across the families (Fig.\ 2), i.e., the variance of the
$p_i$ across families appears to be completely immaterial to the shape
of the distribution---only the average  $\mu \equiv \langle m\rangle$ counts.
\begin{figure}[h]
\centerline{\psfig{figure=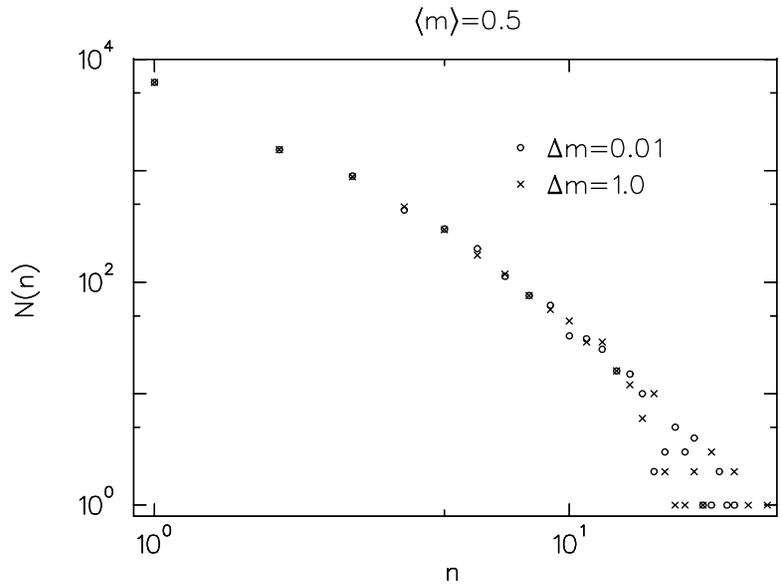,width=4in,angle=90}}
\caption{Abundance patterns obtained from two sets of
numerical simulations of the branching model, each with $\mu=\langle m\rangle =
0.5$. $m$ was chosen from a uniform probability distribution of width
$1$ for the runs represented by crosses, and from a distribution of 
width $0.01$ for those represented by circles.
Simulations where $m$ and $p_i$ are allowed to vary significantly
and those where they are severely constricted are impossible to distinguish
if they share the same $\langle m\rangle$.}
\end{figure}

\begin{figure}[h]
\centerline{\psfig{figure=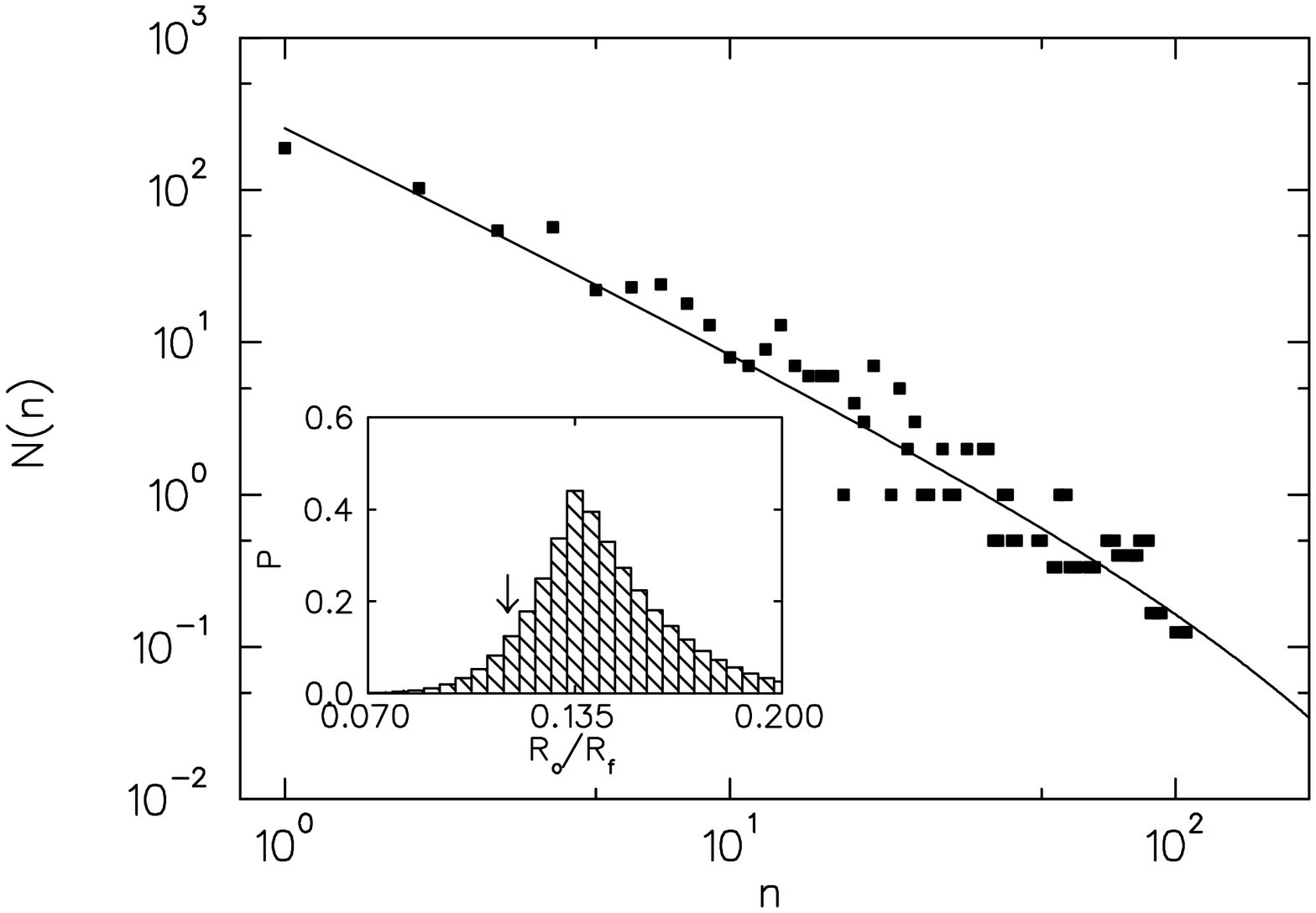,width=4in,angle=90}}
\caption{The abundance distribution of fossil marine animal
orders~\cite{SEPKOSKI} (squares) and the predicted curve from the
branching model (solid line).  The fossil data has been binned above
$n=37$ with a variable bin size~\cite{JCCA2}.  The predicted curve was
generated using $R_o/R_f = N_o/N_f = 0.115$, where $N_o$ and $N_f$
were obtained directly from the fossil data.  The inset shows
Kolmogorov-Smirnov (K-S) significance levels $P$ obtained from
comparison of the fossil data to several predicted distributions with
different values of $R_o/R_f$, which shows that the data is best fit
by $R_o/R_f=0.135$. The arrow points to our prediction $R_o/R_f=0.115$
where $P=0.12$. A Monte Carlo analysis shows that for a sample size of
$626$ (as we have here), the predicted $R_o/R_f=0.115$ is within the
66\% confidence interval of the best fit $R_o/R_f=0.135$ ($P=0.44$).
The K-S tests were done after removal of the first point, which
suffers from sampling uncertainties.}
\end{figure}
In Fig.\ 3, we show the abundance distribution of families within
orders for fossil marine animals~\cite{SEPKOSKI}, together with the
prediction of our branching model.  The theoretical curve was obtained
by assuming that the ratio $R_o/R_f$ is approximated by the ratio of
the total number of orders to the total number of families
\begin{equation}
\frac{R_o}{R_f} \simeq \frac{N_o}{N_f}  \label{HMM}
\end{equation}
and that both are very small compared to the rate of mutations.  The
prediction $\mu=0.9(16)$ obtained from the branching process model by
using (\ref{HMM}) as the sole parameter fits the observed data
remarkably well ($P=0.12$, Kolmogorov-Smirnov test, see inset in
Fig.~3).  Alternatively, we can use a best fit to determine the ratio
$R_o/R_f$ without resorting to (\ref{HMM}), yielding
$R_o/R_f=0.115(20)$ ($P=0.44$). Fitting abundance distributions to the
branching model thus allows us to determine a ratio of parameters
which reflect dynamics intrinsic to the taxon under consideration, and
the niche(s) it inhabits. Indeed, some taxa analyzed in
Refs.~\cite{BURLANDO1,BURLANDO2} are better fit with $0.5<\mu<0.75$,
pointing to conditions in which the rate of taxon formation was much
closer to the rate of subtaxon formation, indicating either a more
``robust'' genome or richer and more diverse niches.

In general, however, Burlando's data~\cite{BURLANDO1,BURLANDO2}
suggest that a wide variety of taxonomic distributions are fit quite
well by power laws ($\mu = 1$).  This seems to imply that actual
taxonomic abundance patterns from the fossil record are characterized
by a relatively narrow range of $\mu$ near $1$. This is likely within
the model description advanced here.  It is obvious that $\mu$ can not
remain above $1$ for significant time scales as this would lead to an
infinite number of subtaxa for each taxon.  What about low $\mu$?  We
propose that low values of $\mu$ are not observed for large (and
therefore statistically important) taxon assemblages for the following
reasons.  If $\mu$ is very small, this implies either a small number
of total individuals for this assemblage, or a very low rate of
beneficial taxon-forming (or niche-filling) mutations. The former
might lead to this assemblage not being recognized at all in field
observations. Either case will lead to an assemblage with too few
taxons to be statistically tractable. Also, since such an assemblage
either contains a small number of individuals or is less suited for
further adaptation or both, it would seem to be susceptible to early
extinction.

The branching model can---with appropriate care---also be applied to
species-abundance distributions, even though these are more
complicated than those for higher taxonomic orders for several
reasons. Among these are the effects of sexual reproduction and the
localized and variable effects of the environment and other species on
specific populations. Still, as the arguments for using a branching
process model essentially rely on mutations which may produce lines of
individuals that displace others, species-abundance distributions may
turn out {\em not} to be qualitatively as different from taxonomically
higher-level rank-frequency distributions as is usually expected.
\begin{figure}[h]
\centerline{\psfig{figure=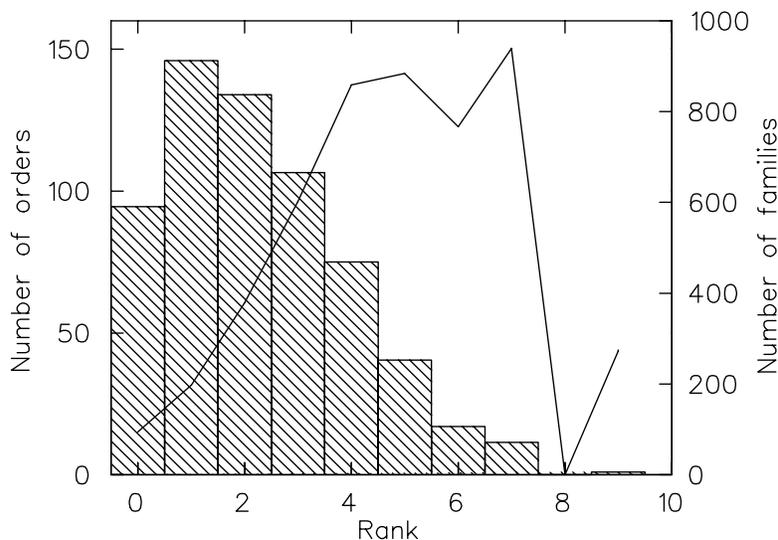,width=4in,angle=90}}
\caption{The abundance distribution of fossil marine
animal orders in logarithmic abundance classes (the same data as Fig.\ 3).
The histogram shows the 
number of orders in each abundance class (left scale), while the solid line 
depicts the number of families in each abundance class (right scale). 
Species rank-abundance distributions where the highest abundance class present
also has the highest number of individuals (as in these data)
are termed {\em canonical lognormal}~\cite{PRESTON2}.}
\end{figure}

Historically, species abundance distributions have been characterized 
using frequency histograms of the number of species in logarithmic 
abundance classes.  For many taxonomic assemblages, this was found 
to produce a humped distribution truncated on the left---a shape usually 
dubbed {\em lognormal}~\cite{PRESTON1,PRESTON2,SUGIHARA}.  
In fact, this distribution is
not incompatible with the power-law type distributions described
above. Indeed, plotting the fossil data of Fig.\ 3 in logarithmic
abundance classes produces a lognormal (Fig.\ 4). 

For species, $\mu$ is the mean number of children each individual of
the species has. (Of course, for sexual species, $\mu$ would be half
the mean number of children per individual.)  In the present case,
$\mu$ less than $1$ implies that extant species' populations {\em
  decrease} on average, while $\mu$ equal to $1$ implies that average
populations do not change.  An extant species' population can decline
due to the introduction of competitors and/or the decrease of the size
of the species' ecological niche.  Let us examine the former more
closely. If a competitor is introduced into a saturated niche, all
species currently occupying that niche would temporarily see a
decrease in their $m$ until a new equilibrium was obtained. If the new
species is significantly fitter than the previously existing species,
it may eliminate the others. If the new species is significantly less
fit, then it may be the one eliminated. If the competitors are about
as efficient as the species already present, then the outcome is less
certain. Indeed, it is analogous to a non-biased random walk with a
possibility of ruin.  The effects of introducing a single competitor
are transient. However, if new competitors are introduced more or less
periodically, then this would act to push $m$ lower for all species in
this niche and we would expect an abundance pattern closer to the
exponential curve as opposed to the power-law than otherwise expected.
We have examined this in simulations of populations where new
competitors were introduced into the population by means of neutral
mutations---mutations leading to new species of the same fitness as
extant species---and found that these are fit very well by the
branching model.  A higher rate of neutral mutations and thus of new
competitors leads to distributions closer to exponential.  We have
performed the same experiment in more sophisticated systems of digital
organisms (artificial life)~\cite{AVIDA,SANDA} and found the same
result~\cite{JCCA2}.

If no new competitors are introduced but the size of the niche
is gradually reduced, we expect the same effect on $m$ and on the
abundance distributions. Whether it is possible to separate
the effects of these two mechanisms in ecological abundance patterns
obtained from field data is an open question.
An analysis of such data to examine these trends would certainly 
be very interesting.

So far, we have sidestepped the difference between historical and
ecological distributions.  For the fossil record, the historical
distribution we have modeled here should work well. For field
observations where only currently living groups are considered, the
nature of the death and extinction processes for each group will
affect the abundance pattern.  In our simulations and artificial-life
experiments, we have universally observed a strong correlation between
the shapes of historical and ecological distributions.  We believe
this correspondence will hold in natural distributions as well when
death rates are affected mainly by competition for resources.
The model's validity for different scenarios is an interesting question,
which could be answered by comparison with more taxonomical data.

Our branching process model allows us to reexamine 
the question of whether any type of special
dynamics---such as self-organized criticality~\cite{soc} (SOC)---is 
at work in evolution~\cite{BakSneppen,Adami-soc}. While showing that the
statistics of taxon rank-frequency patterns in evolution are
closely related to the avalanche sizes in SOC sandpile models,
the present model clearly shows that
instead of a subsidiary relationship where evolutionary processes may
be self-organized critical, the power-law
behaviour of both evolutionary {\em and} sandpile
distributions can be understood in terms of the mechanics
of a Galton-Watson branching process~\cite{JCCA2,ZAPPERI}. 
The mechanics of this branching process 
are such that the branching trees are probabilistic
fractal constructs. However, the underlying stochastic process
responsible for the observed behaviour can be
explained simply in terms of a random walk~\cite{SPITZER}. 
For evolution, the propensity for near power-law behaviour is found to 
stem from a dynamical process in which $\mu\approx1$ is selected for and
highly more likely to be observed than other values, while the
``self-tuning'' of the SOC models is seen to result from arbitrarily 
enforcing conditions which would correspond to the limit
$R_o/R_f \rightarrow 0$ and therefore
$m \rightarrow 1$~\cite{JCCA2}.
\newpage

\bibliographystyle{unsrt}

\vskip 0.25in

\noindent {\bf Acknowledgments.} We would like to thank J. J. Sepkoski
for kindly sending us his amended data set of fossil marine animal families.
Access to the Intel Paragon XP/S was provided by the Center of
Advanced Computing Research at the California Institute of Technology.
This work was supported by a grant from the NSF.
\vskip 0.25cm

\noindent Correspondence and requests for materials should be addressed to C.A.
(e-mail: adami@krl.caltech.edu).

\end{document}